\documentclass[floats,prb,twocolumn,amsmath,superscriptaddress]{revtex4}

\usepackage{graphicx}
\usepackage{color}
\usepackage{dcolumn}

\begin{document}

\newcommand{\ket}    [1]{{|#1\rangle}}
\newcommand{\bra}    [1]{{\langle#1|}}
\newcommand{\braket} [2]{{\langle#1|#2\rangle}}
\newcommand{\bracket}[3]{{\langle#1|#2|#3\rangle}}
\newcommand{\red}{\textcolor{red}}
\newcommand{\blue}{\textcolor{blue}}
\newcommand{\green}{\textcolor{green}}
\newcommand{\cyan}{\textcolor{cyan}}
\newcommand{\magenta}{\textcolor{magenta}}

\def\pw{^{({\rm W})}}
\def\ph{^{({\rm H})}}
\def\k{{\bf k}}
\def\R{{\bf R}}
\def\b{{\bf b}}
\def\q{{\bf q}}
\def\o{{\cal O}}
\def\e{{\cal E}}
\def\v{{\rm v}}
\def\pw{^{({\rm W})}}
\def\ph{^{({\rm H})}}
\def\la{\langle\kern-2.0pt\langle}
\def\ra{\rangle\kern-2.0pt\rangle}
\def\vt{\vert\kern-1.0pt\vert}
\def\D{{D}\ph}
\def\n{{\cal N}}
\def\u{{\cal U}}

\title{Treatment of a semi-metal to metal
structural phase transition: 

convergence properties
of the A7~$\to$~sc transition of arsenic
}

\author{Patricia Silas}
\affiliation{Theory of Condensed Matter, Cavendish Laboratory,
University of Cambridge, JJ Thomson Avenue, Cambridge CB3 0HE, United Kingdom}

\author{Jonathan R. Yates}
\affiliation{Theory of Condensed Matter, Cavendish Laboratory,
University of Cambridge, JJ Thomson Avenue, Cambridge CB3 0HE, United Kingdom}

\author{Peter D. Haynes}
\affiliation{Departments of Materials and Physics,
Imperial College London, Exhibition Road, London SW7 2AZ, United Kingdom}

\begin{abstract}
\vspace{0.4mm}
The material presented here is supplementary to the article entitled:
``Density-functional investigation of the rhombohedral to simple cubic phase
transition of arsenic''; it deals with the convergence
issues involved in studying a semi-metal to metal structural
phase transition such as the A7~$\to$~sc transition of arsenic.
\end{abstract}

\maketitle

The occurrence of the {A}7~$\to$~sc phase transition of arsenic
is identified most clearly by the behavior of the nearest and next-nearest neighbor distances.
In this study we investigate closely the convergence of these distances with respect
to \mbox{$k$-point} grid size and smearing over the pressure ranges
of \mbox{12--24}~GPa and \mbox{17--30}~GPa for the {LDA} and {GGA-PBE} cases, respectively.
In the region of the transition, the energy differences between
the two structures are extremely small.  If for a given pressure
the potential energy surface is noisy and rather flat, then it is possible
that our system relaxes to a structure that corresponds to a local
minimum, rather than the global minimum of our energy surface.
We have found that to be able to achieve good results we must 
resolve our potential energy surface as much as possible.
We have investigated Gaussian smearing,\cite{fu_83} and Methfessel-Paxton smearing,\cite{methfessel_89}
but we find that we obtain much cleaner and clearer results by
using cold-smearing \cite{marzari_99} and therefore have
done so throughout this study (a review of these different smearing techniques 
can be found in Ref.~\onlinecite{walker_04}).

In these numerical calculations, the integrations that are to be
performed over the {B}rillouin zone
are discretized by way of the \mbox{$k$-point} grid.
In the case of an insulator, no smearing is required.
Smearing is only required in the event that one or more bands cross the Fermi level,
in other words if the material possesses a Fermi surface (if it is a metal or a
semi-metal).
The effect of smearing is to blur the details of the Fermi surface 
(by imposing an artificial temperature on the electronic system);
smearing must be used to ensure that the calculation is stable,
but can also be used to hasten the convergence of the calculation.
If our material does possess a Fermi surface, but we use an infinitely dense
\mbox{$k$-point} grid, no smearing is needed as no details of the
Fermi surface will be lost;
the choice of \mbox{$k$-point} grid and smearing are coupled to each other.
We would like to minimize the number of \mbox{$k$-points} used in the calculation,
but must choose a grid dense enough to properly sample the 
{B}rillouin zone as well as to pick out the features of the Fermi surface.
If we use a smaller (less dense) \mbox{$k$-point} grid, we must use more smearing, 
but the exact amount of smearing to use is not obvious.  If we use a 
smearing that is too high, details of the Fermi surface may be
washed out possibly to the extent of affecting the quantities that
interest us.  It is preferable to keep the smearing low to obtain
results of the highest accuracy possible, but there is a limit
to how low the smearing can be set for a given \mbox{$k$-point} grid
before there is a risk that the calculation becomes unstable.

In fact, to choose the smearing properly
for a particular \mbox{$k$-point} grid would require knowledge
of the three-dimensional band-structure of the material throughout
the entire {B}rillouin zone, since the smearing should be determined
by the steepest gradients of the bands crossing the Fermi level.
(Although we cannot avoid it, it is not actually appropriate to apply a unique
amount of smearing to the entire band-structure of a material
undergoing a semi-metal to metal phase transition;
ideally we would use some sort of adaptive smearing technique to
minimize the washing-out of the features of the Fermi surface.)
It is not feasible to investigate three-dimensional
band-structures, so we must carefully converge our 
calculations with respect to both \mbox{$k$-point} grid size
and smearing.  Initially, we did some cursory convergence tests
to determine suitable choices for the cut-off energy as well as for the
density of the fine-grid, respectively to ensure an appropriate
basis set size and to recover sufficient detail of the charge
density within the atomic cores; these tests were performed
on the uncompressed system.  The results of our geometry optimizations, however, must
be subjected to more rigorous convergence testing; each calculation
must be repeated in its entirety for
each combination of \mbox{$k$-point} grid and smearing.
The properties that interest us must be converged in this way
before anything can reliably be said about them.
Although convergence tests for \mbox{$k$-point}
grid size are standard,
in electronic structure calculations that involve smearing
it is very rare to find that testing
has been undertaken to determine
how the smearing has affected the results;
yet applying a casual approach to the choice of the amount of smearing used
can lead to gross inaccuracies (see Ref.~\onlinecite{mehl_00}). 
Thus, it is essential that tests for convergence with 
respect to both $k$-point grid and smearing be performed.
Furthermore, it must be stressed that such tests should not merely be carried out
by looking at what happens to the total energy of the system; rather,
these tests must
always be carried out on the quantities specific to the study itself.

\begin{figure*}
\begin{center}
\includegraphics[width=0.85\textwidth]{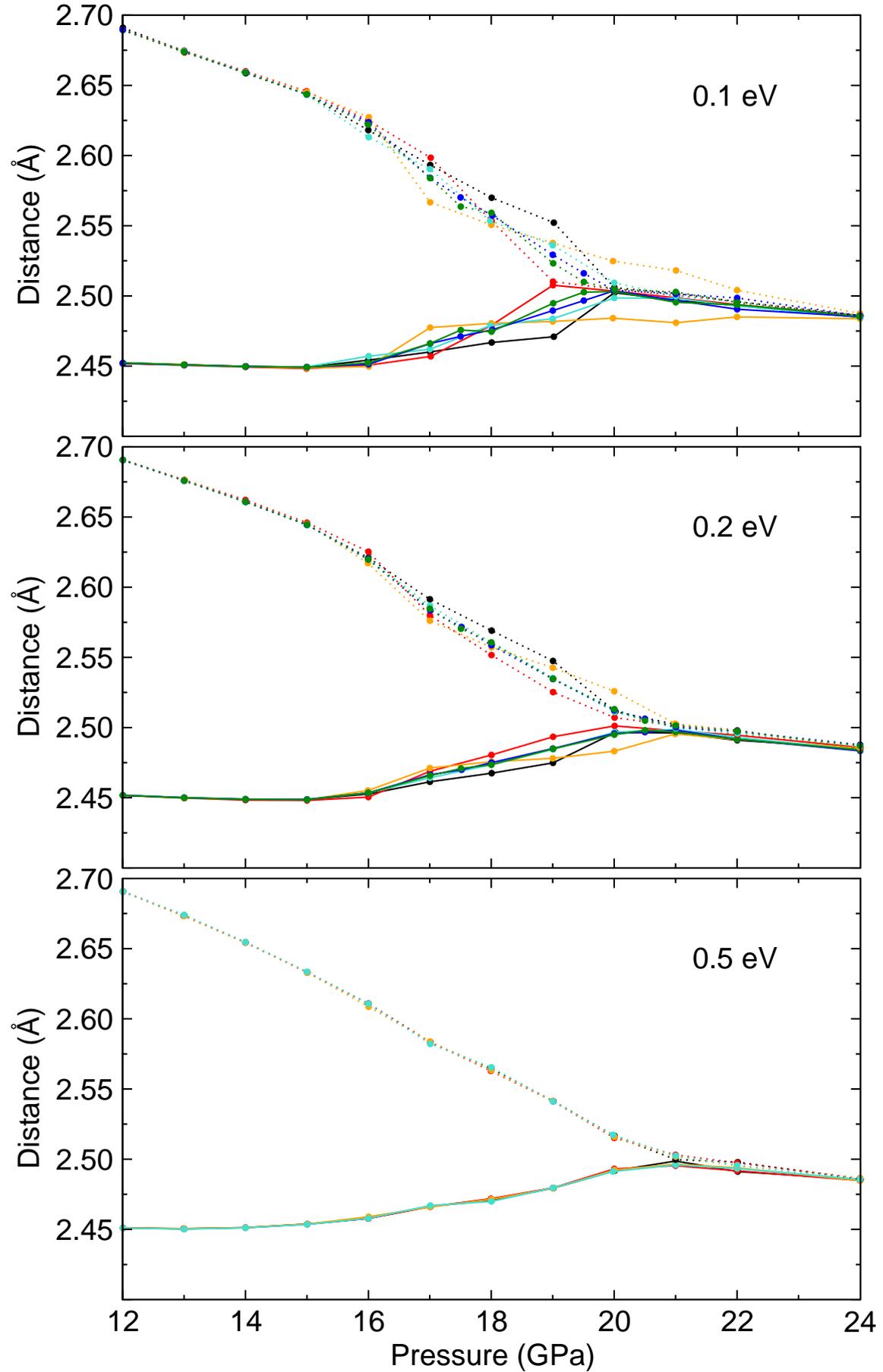}
\caption{Nearest neighbor (solid lines) and next-nearest neighbor
(dotted lines) distances as a function of pressure
for the {A}7~$\to$~sc transition of arsenic using the LDA. Grids used are: 
24$\times$24$\times$24 (black), 25$\times$25$\times$25 (red), 26$\times$26$\times$26 (orange),
33$\times$33$\times$33 (cyan), 50$\times$50$\times$50 (blue) and 66$\times$66$\times$66 (green).
The values of cold-smearing used are: 0.1~eV (top), 0.2~eV (middle) and 0.5~eV (bottom).}
\label{fig: nn_and_nnn_LDA}
\end{center}
\end{figure*}

As we mentioned above, we would like to perform an in-depth
investigation of the {A}7~$\to$~sc phase transition and so we must
test the convergence of our quantities of interest,
the nearest and next-nearest neighbor distances,
with respect to $k$-point grid and smearing in the
region of the transition.
Fig.~\ref{fig: nn_and_nnn_LDA} displays the results of these convergence
tests for the case of the {LDA}.  Consider first the top panel of this 
figure, which displays the behavior of the nearest and next-nearest
neighbor distances for different grid sizes using a cold-smearing
of 0.1~eV.  We see that for this value of the smearing, none of the
results for the different grid sizes have converged in the immediate vicinity of
the transition, which appears to occur somewhere in the range of \mbox{19--21}~GPa.
Even our results for a 50$\times$50$\times$50 grid have
not quite converged to those for a 66$\times$66$\times$66 grid. Note also
how different our results at the transition are for the
24$\times$24$\times$24, 25$\times$25$\times$25
and 26$\times$26$\times$26 grids at this smearing. 
This is down to the particular sampling of the {B}rillouin zone
for each specific grid and tells us that we are not using enough
\mbox{$k$-points} to properly sample the {B}rillouin zone for this smearing and at these pressures.
Outside the region of the phase transition, the calculations are
less sensitive to the density of the \mbox{$k$-point} grid;
below about 14~GPa and above about 30~GPa we see that our results
are well converged at this smearing for all grid sizes.  Hence we can
be confident about our choice of grid and smearing (33$\times$33$\times$33,
 0.1~eV) at pressures away from this particular transition.

If we decide that we do not require the finest features of the Fermi surface
to obtain a meaningful result, then we can increase the smearing, as we have done 
for the middle panel of Fig.~\ref{fig: nn_and_nnn_LDA}, for which a cold-smearing
of 0.2~eV was used. We see that at this smearing, the results have converged
for the three most dense grids used, 33$\times$33$\times$33, 50$\times$50$\times$50
and 66$\times$66$\times$66, and the transition pressure seems to have shifted
slightly higher appearing now to occur at \mbox{20--21}~GPa.

In the lowest panel of Fig.~\ref{fig: nn_and_nnn_LDA}, we have increased
the smearing high enough to ensure that the results of all of our grids
converge.  Having shown, for a smearing of 0.5~eV,
the convergence of the calculations for the least
dense grids used, we do not repeat the calculations for the most dense
grids of 50$\times$50$\times$50 and 66$\times$66$\times$66. 
If we compare the top and bottom panels,
we can see quite clearly that as the smearing is increased, the
transition shifts to slightly higher pressures.

\begin{figure*}
\begin{center}
\includegraphics[width=0.85\textwidth]{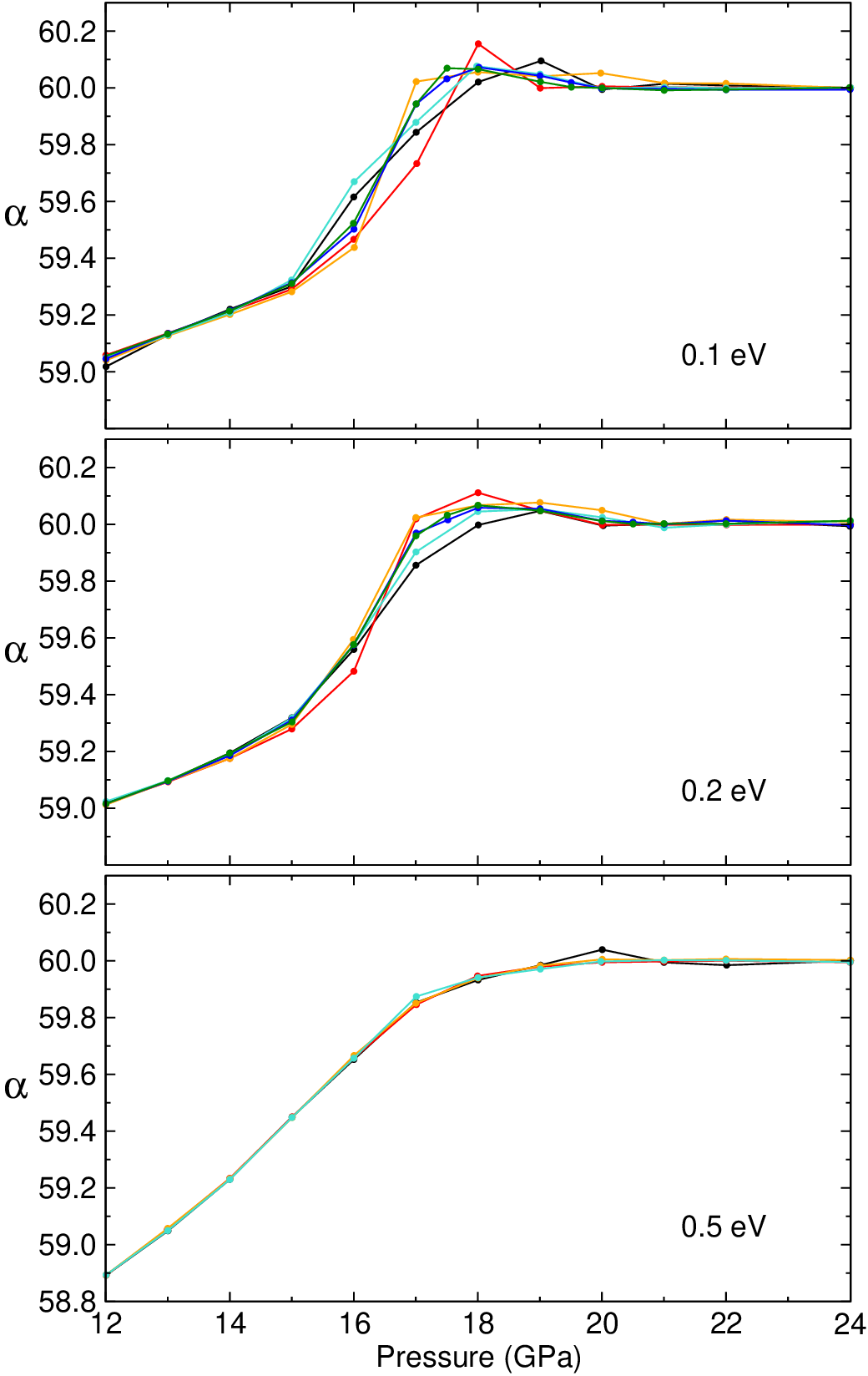}
\caption{Cell angle $\alpha$ as a function of pressure
for the {A}7~$\to$~sc transition of arsenic using the LDA. Grids used are:
24$\times$24$\times$24 (black), 25$\times$25$\times$25 (red), 26$\times$26$\times$26 (orange),
33$\times$33$\times$33 (cyan), 50$\times$50$\times$50 (blue) and 66$\times$66$\times$66 (green).
The values of cold-smearing used are: 0.1~eV (top), 0.2~eV (middle) and 0.5~eV (bottom).
}
\label{fig: alpha_LDA}
\end{center}
\end{figure*}

\begin{figure*}
\begin{center}
\includegraphics[width=0.85\textwidth]{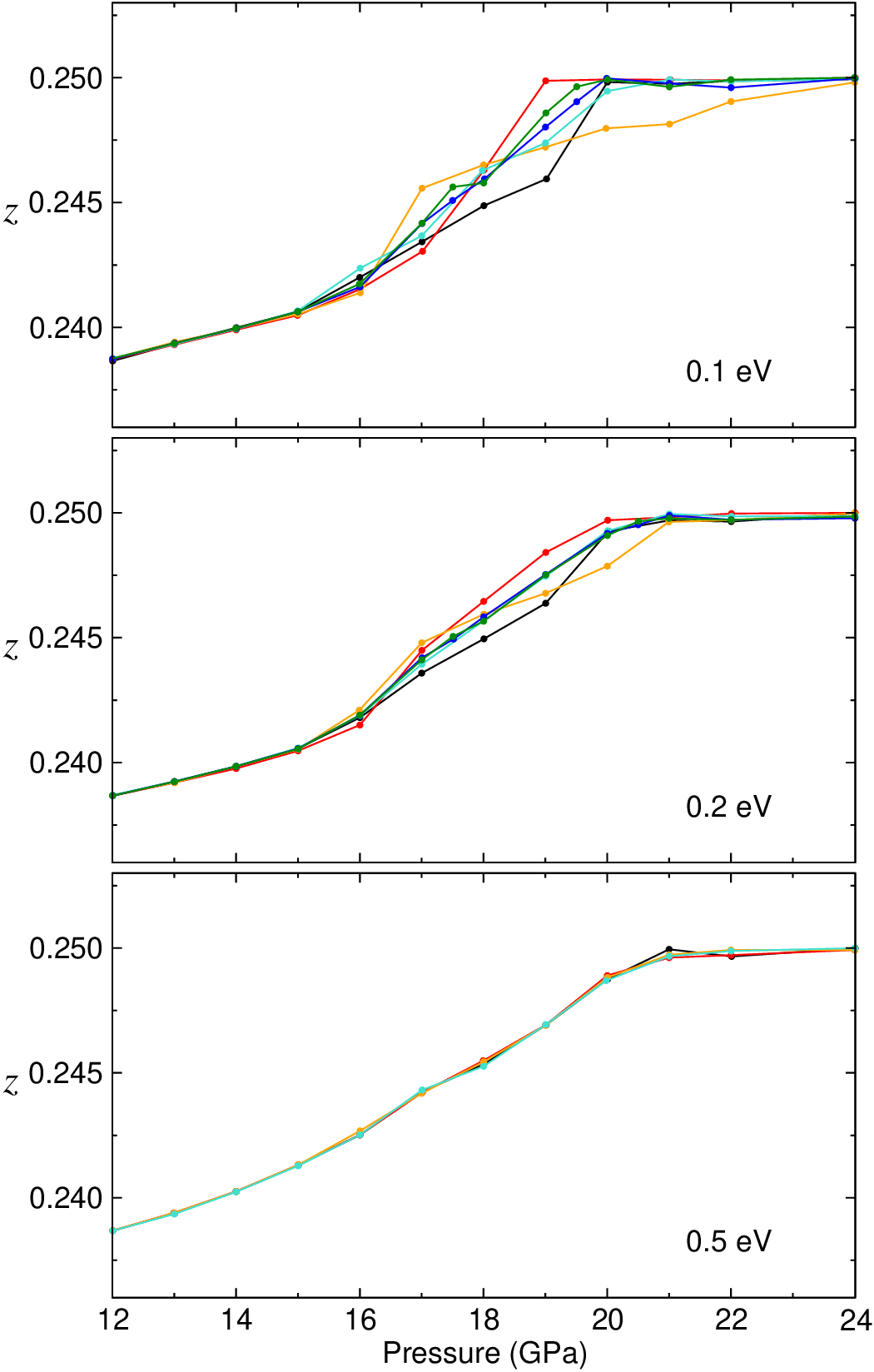}
\caption{Atomic positional parameter $z$ as a function of pressure
for the {A}7~$\to$~sc transition of arsenic using the LDA. Grids used are:
24$\times$24$\times$24 (black), 25$\times$25$\times$25 (red), 26$\times$26$\times$26 (orange),
33$\times$33$\times$33 (cyan), 50$\times$50$\times$50 (blue) and 66$\times$66$\times$66 (green).
The values of cold-smearing used are: 0.1~eV (top), 0.2~eV (middle) and 0.5~eV (bottom).
}
\label{fig: z_LDA}
\end{center}
\end{figure*}

\begin{figure*}
\begin{center}
\includegraphics[width=0.85\textwidth]{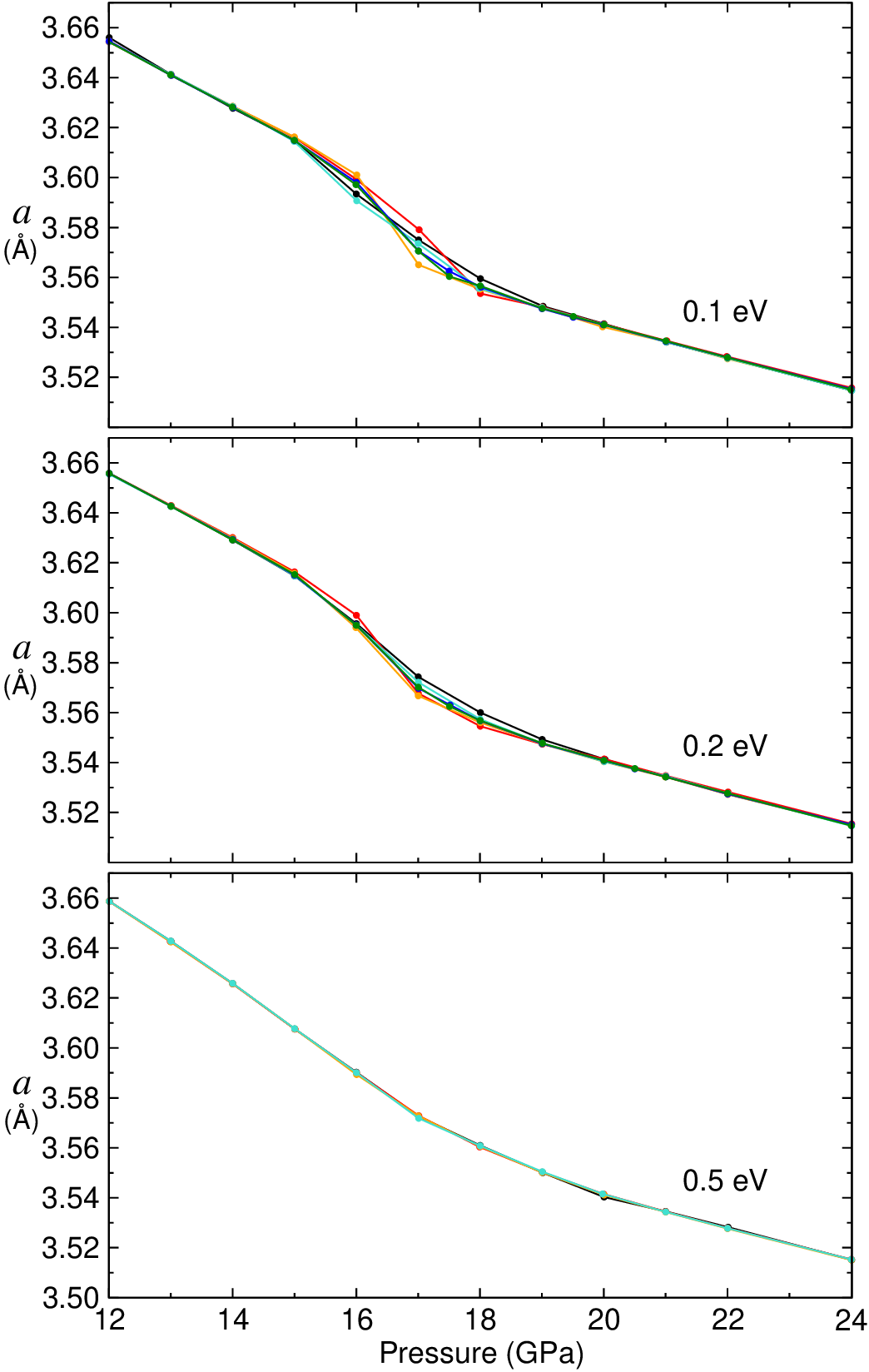}
\caption{Lattice parameter $a$ as a function of pressure
for the {A}7~$\to$~sc transition of arsenic using the LDA. Grids used are:
24$\times$24$\times$24 (black), 25$\times$25$\times$25 (red), 26$\times$26$\times$26 (orange),
33$\times$33$\times$33 (cyan), 50$\times$50$\times$50 (blue) and 66$\times$66$\times$66 (green).
The values of cold-smearing used are: 0.1~eV (top), 0.2~eV (middle) and 0.5~eV (bottom).
}
\label{fig: a_LDA}
\end{center}
\end{figure*}

The components that make up the nearest and next-nearest neighbor
distances are the cell angle $\alpha$, the atomic positional parameter $z$,
and the lattice parameter $a$.  We can therefore
examine the convergence properties
across the transition of these constituent quantities as well.
In the case of the LDA, the results of our convergence testing
on  $\alpha$, $z$, and $a$ are presented in Fig.~\ref{fig: alpha_LDA},
Fig.~\ref{fig: z_LDA} and Fig.~\ref{fig: a_LDA}, respectively.

\begin{figure*}
\begin{center}
\includegraphics[width=0.85\textwidth]{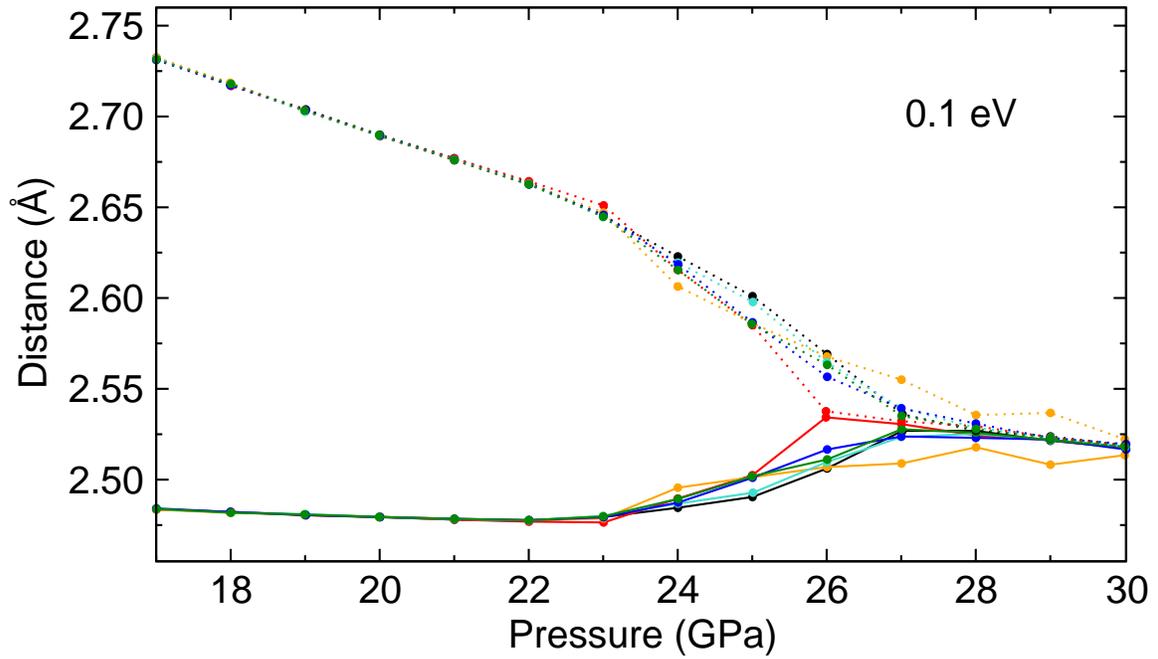}
\caption{Nearest neighbor (solid lines) and next-nearest neighbor
(dotted lines) distances as a function of pressure
for the {A}7~$\to$~sc transition of arsenic using the GGA-PBE. Grids used are: 
24$\times$24$\times$24 (black), 25$\times$25$\times$25 (red), 26$\times$26$\times$26 (orange),
33$\times$33$\times$33 (cyan), 50$\times$50$\times$50 (blue) and 66$\times$66$\times$66 (green),
each with a cold-smearing of 0.1~eV.}
\label{fig: nn_and_nnn_PBE}
\end{center}
\end{figure*}

\begin{figure*}
\begin{center}
\includegraphics[width=0.85\textwidth]{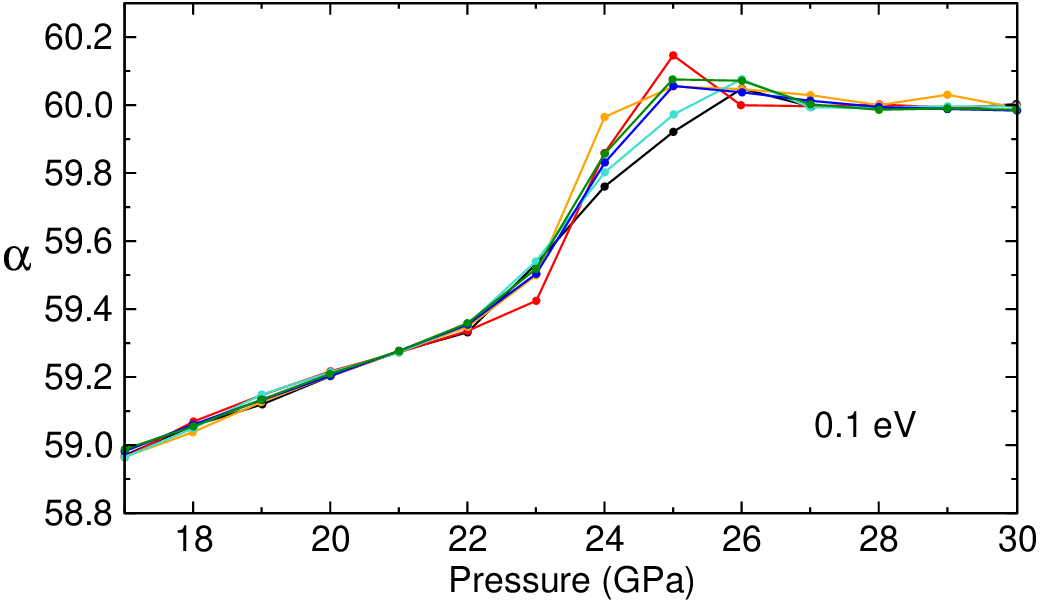}
\caption{Cell angle $\alpha$ as a function of pressure
for the {A}7~$\to$~sc transition of arsenic using the GGA-PBE. Grids used are:
24$\times$24$\times$24 (black), 25$\times$25$\times$25 (red), 26$\times$26$\times$26 (orange),
33$\times$33$\times$33 (cyan), 50$\times$50$\times$50 (blue) and 66$\times$66$\times$66 (green),
each with a cold-smearing of 0.1~eV.
}
\label{fig: alpha_PBE}
\end{center}
\end{figure*}

\begin{figure*}
\begin{center}
\includegraphics[width=0.85\textwidth]{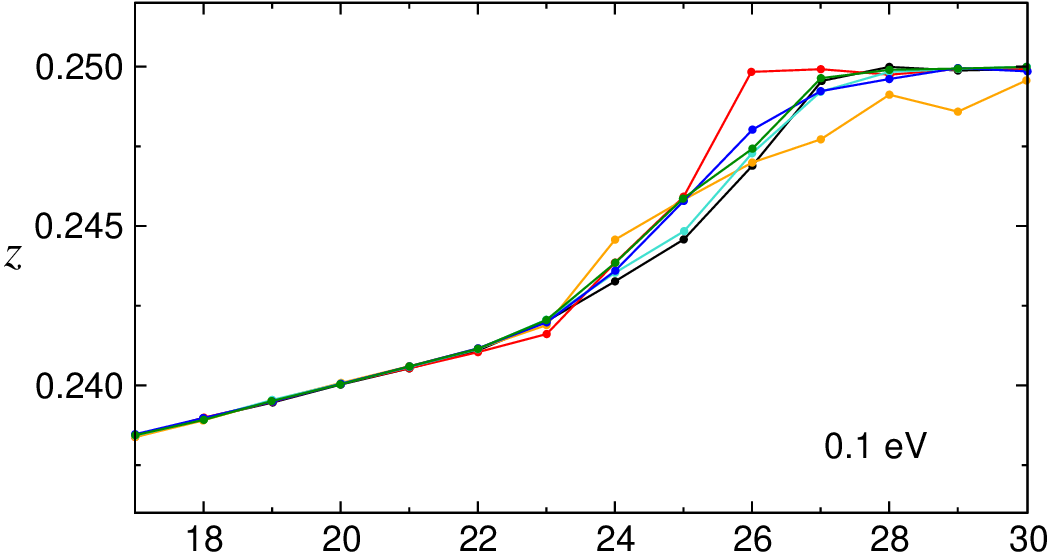}
\caption{Atomic positional parameter $z$ as a function of pressure
for the {A}7~$\to$~sc transition of arsenic using the GGA-PBE. Grids used are:
24$\times$24$\times$24 (black), 25$\times$25$\times$25 (red), 26$\times$26$\times$26 (orange),
33$\times$33$\times$33 (cyan), 50$\times$50$\times$50 (blue) and 66$\times$66$\times$66 (green),
each with a cold-smearing of 0.1~eV.
}
\label{fig: z_PBE}
\end{center}
\end{figure*}

\begin{figure*}
\begin{center}
\includegraphics[width=0.85\textwidth]{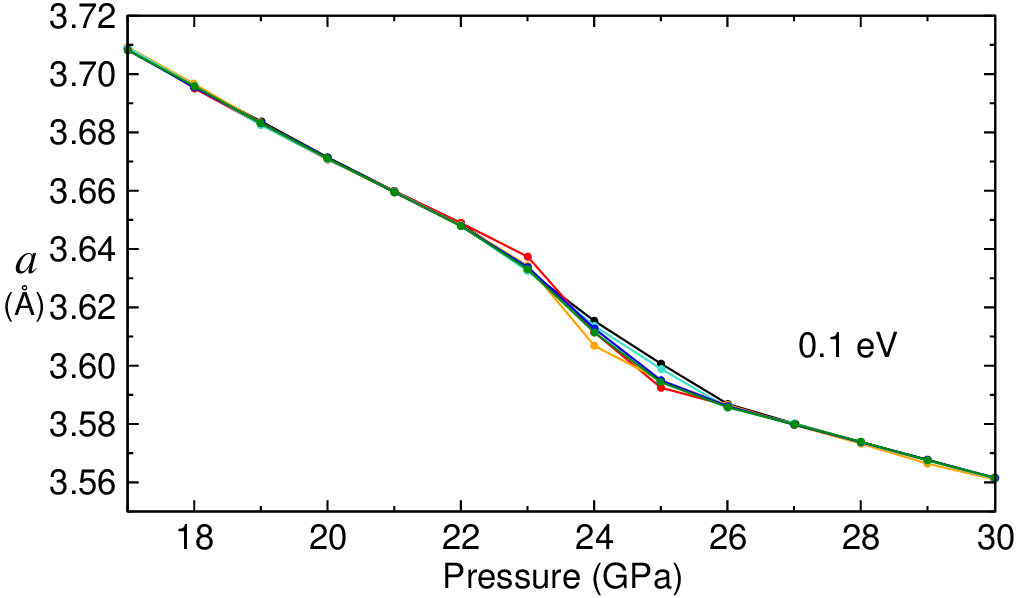}
\caption{Lattice parameter $a$ as a function of pressure
for the {A}7~$\to$~sc transition of arsenic using the GGA-PBE. Grids used are:
24$\times$24$\times$24 (black), 25$\times$25$\times$25 (red), 26$\times$26$\times$26 (orange),
33$\times$33$\times$33 (cyan), 50$\times$50$\times$50 (blue) and 66$\times$66$\times$66 (green),
each with a cold-smearing of 0.1~eV.
}
\label{fig: a_PBE}
\end{center}
\end{figure*}

We observe similar difficulty in the convergence of the {A}7~$\to$~sc phase
transition in the case of the {GGA-PBE}.  Fig.~\ref{fig: nn_and_nnn_PBE} shows
the results in terms of the nearest and next-nearest neighbor distances
of our {GGA-PBE} calculations for each of the same grids employed
for the {LDA}, but for a cold-smearing of 0.1~eV only.  Considering the
data resulting from the use of the most dense \mbox{$k$-point} grids, we can
conclude from this figure that in this case the phase transition 
must happen over the pressure interval of \mbox{27--29}~GPa.  The 
corresponding results of the convergence testing on $\alpha$, $z$, and $a$
in this case are revealed in Fig.~\ref{fig: alpha_PBE},
Fig.~\ref{fig: z_PBE} and Fig.~\ref{fig: a_PBE}, respectively.

As we said above, for a cold-smearing of 0.1~eV even calculations using
a 50$\times$50$\times$50 \mbox{$k$-point} grid have not converged to those
using a 66$\times$66$\times$66 grid in the vicinity of the transition.
We can conclude just by observing the top panel of Fig.~\ref{fig: nn_and_nnn_LDA}
that in the vicinity of the phase transition, 
these calculations are extremely sensitive to the details of the Fermi
surface.  The Fermi surface of arsenic is extremely complex.  From work carried out
by Lin and Falicov,\cite{linandfalicov_66} we estimate that a \mbox{$k$-point} grid
at least twice as dense as our most dense grid used (66$\times$66$\times$66) would
be required to resolve all of the features of the Fermi surface of arsenic at
ambient pressures.

The most dense grid employed by \mbox{Da Silva, \textit{et. al.}\cite{dasilva_97}} 
in their study of arsenic was 13$\times$13$\times$13, and in the case of
H{\"{a}}ussermann,~\textit{et. al.}~\cite{haussermann_02} it was 17$\times$17$\times$17.
Durandurdu~\cite{durandurdu_05} uses only the gamma point for a unit cell containing
250 atoms, roughly corresponding to using a 5$\times$5$\times$5 \mbox{$k$-point} grid for
for a two-atom unit cell.  Our investigations reveal that these calculations could not
have been converged, and that any agreement with experiment that
may have been observed in these cases was merely fortuitous.

We conclude that it is surprisingly difficult to converge the calculations
for this semi-metal to metal phase transition using reasonably
sized $k$-point grids and smearings.  
It is of course possible to converge the calculations using less dense grids
if very large smearings are used  
but it would be at the expense of accuracy in the resulting transition pressures.
To ensure accuracy when studying a pressure-induced semi-metal to metal phase
transition, dense \mbox{$k$-point} grids are essential.

\bibliographystyle{apsrev}

\end{document}